# Facial Expression Cloning with Elastic and Muscle Models

Yihao Zhang[1], Weiyao Lin[1*], Bing Zhou[2], Zhenzhong Chen[3], Bin Sheng[1], Jianxin Wu[4], and Wenjun Zhang[1]

[1] Department of Electronic Engineering, Shanghai Jiao Tong University, Shanghai, China

[2] School of Information Engineering, Zhengzhou University, China

[3] Wuhan University, China

[4] Department of Computer Science, Nanjing University, Nanjing 210023, China

*Corresponding author, Email: hellomikelin@gmail.com, Phone: 86-21-34208843, Fax: 86-21-34204155

*Abstract*—Expression cloning plays an important role in facial expression synthesis. In this paper, a novel algorithm is proposed for facial expression cloning. The proposed algorithm first introduces a new elastic model to balance the global and local warping effects, such that the impacts from facial feature diversity among people can be minimized, and thus more effective geometric warping results can be achieved. Furthermore, a muscle-distribution-based (MD) model is proposed, which utilizes the muscle distribution of the human face and results in more accurate facial illumination details. In addition, we also propose a new distance-based metric to automatically select the optimal parameters such that the global and local warping effects in the elastic model can be suitably balanced. Experimental results show that our proposed algorithm outperforms the existing methods.

**Keywords:** Facial expression cloning; Elastic model; Muscle-distribution-based model; Distance-based metric

## I. Introduction and Related Works

Facial expression synthesis is of increasing importance in many applications such as movie making, video conferencing, and video games [1-8, 12-18]. In this area, expression cloning (or mapping) is one of the most effective techniques for synthesizing facial expressions [1-6]. Basically, the target for facial expression cloning is to transfer one person's expression (i.e., the source person) to another person's neutral face (i.e., the target person), thus the second person's facial expression can be synthesized, as shown in Fig. 1.

Several algorithms have been proposed for expression cloning [1-10,16]. Most algorithms utilize geometry warping [1-6,16] or motion cloning [7-8] on the face feature positions or the triangulated meshes to map facial expressions. For example, Sumner and Popovic [8] utilize nonlinear deformation transfer to map the 3D motions or expressions from one source object to the target object. Song et al. [2] use vertex tent coordinate transfer to perform geometric warping based on 3D models. Seol and Lewis [15] utilize Radial Basis Functions and movement matching to determine the warping. However, since these methods perform warping globally while the local facial feature differences between people (e.g., face shape differences, mouth or eye differences) are not well considered, their facial expression cloning results are less satisfactory in some cases. Noh and Neumann [7] aim to refine the local parts of the face based on the muscle model of human face [10]. Although they can create good facial feature motions for the same person, their methods are less effective when applied to synthesize the expression of another person as the facial feature differences between people are still neglected.

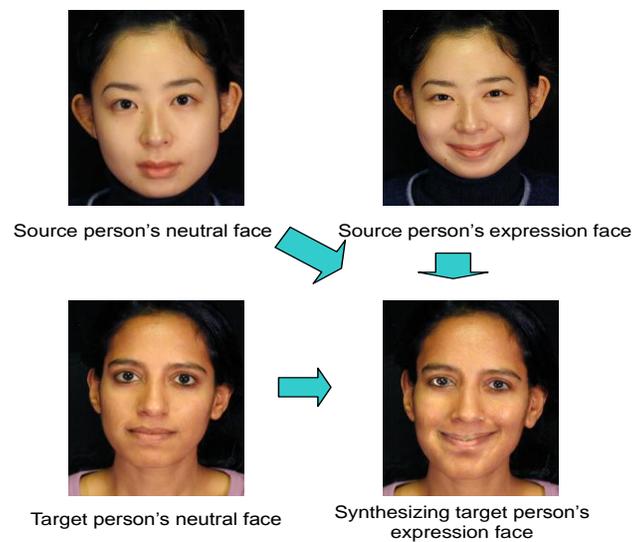

Fig. 1 The process for facial expression cloning.

Besides, since the facial illumination details will also change in different expressions, this detailed illumination information also needs to be transferred to the target person for creating more vivid expression results. However, most existing algorithms only focus on the geometry warping of the face while the illumination details are ignored. Although some researchers introduced expression ratio image (ERI) [5] or mesh image [2] to transfer the illumination details, their methods still have limitations due to face feature differences, unsuitable noise filtering, or detail importance differences.

In this paper, a new elastic-plus-muscle-distribution-based (E+MD) algorithm is proposed for facial expression cloning. Our contribution can be summarized as follows: (1) A new elastic model is proposed to balance the global and local warping effects such that the effects from facial feature differences between people can be avoided. Thus more reasonable geometric warping results can be created. (2) A muscle-distribution-based (MD) model is proposed which utilizes the muscle distribution information of the human face to evaluate and strengthen the facial illumination details. By this way, the effects from human face difference as well as the impacts of unsuitable noise filtering can be effectively reduced. (3) A new distance-based (DB) metric is designed to automatically select the parameters in the elastic model. The proposed DB metric evaluates the facial expression cloning results based on the expression distance among the synthesized target person's expression face, the source person's expression face, and the target person's globally warped face. With the proposed DB metric, the global and local warping effects in the elastic model can be suitably balanced. Experimental results demonstrate the effectiveness of our proposed method.

The rest of the paper is organized as follows: Section II describes the motivations of our proposed E+MD algorithm. Section III describes the details of our algorithm. Section IV shows the experimental results. Section V concludes the paper.

## II. MOTIVATION OF THE PROPOSED ALGORITHM

As aforementioned, geometric warping can be used for creating the target person's expression face [1-6]. In the geometric warping process, the face feature positions are first identified for each face (e.g., Fig. 2 (a)). Many algorithms can be used to automatically achieve these feature positions such as the active appearance model (AAM) [12]. In the experiments of this paper, we manually identify the feature points in order to have fair comparisons with the other methods [2,5]. However, similar results can be achieved when automatic feature position identification methods [12,15] are used in our experiments.

Then, triangulation can be performed for creating the triangle meshes according to these feature positions (Fig. 2 (b)). In this paper, the Delaunay triangulation method is used to get the triangle meshes [5]. Based on the triangulated mesh information, the result of the target expression face can be achieved by geometric-controlled image warping (i.e., warp the pixels of the target person's neutral face to create the target person's expression face based on the corresponding warp ratio between the pixels in the source person's neutral face and the source person's expression face) [3,5]. Fig. 2 (c) shows one geometric warping result from the three available images in Fig. 1 [3,5] (i.e., the source person's neutral face, source person's expression face, and target person's neutral face images, as in Fig. 1). However, the warping result in Fig. 2 (c) is unsatisfactory since it looks not so much like the expression "smile". This is because most of the existing geometric warping algorithms [1-6,8] are performed "globally" where the feature positions in the target expression image are moved "relatively" according to the movements of their corresponding feature positions in the source person's face. Since people may often have very different local features on their faces, these "relative" movements may often fail to create vivid expression results. For example, consider a source person whose mouth is small in the neutral face and whose mouth becomes much larger in the "smile" expression face. If we apply this smile expression to a target person whose mouth is already large in the neutral face, the global warping method may lead to an unnatural "super large" mouth. Although some registration steps can be applied to scale the expression changes [3-8], such problem still cannot be avoided.

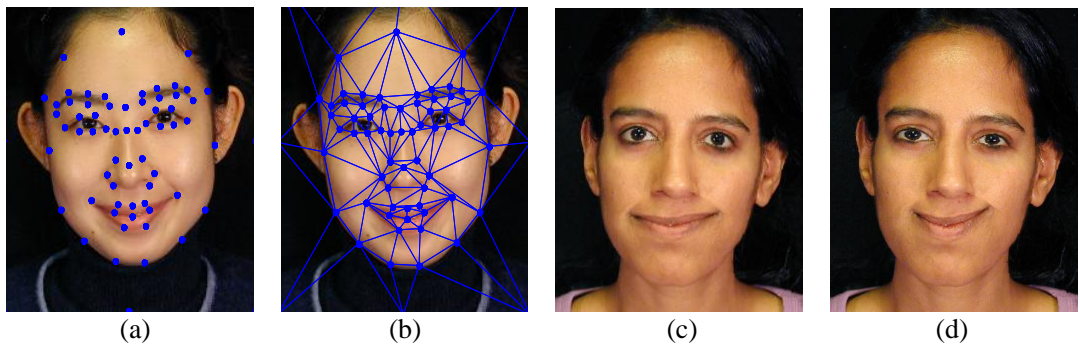

(a) (b) (c) (d)

Fig. 2 (a): The identified face feature positions of the source person's expression face in Fig. 1, (b): The triangulated meshes according to (a), (c): The "global" geometric warped result from the three available images in Fig. 1 [3], (d): The "local" geometric warped result from the available images in Fig. 1.

In order to overcome the problem of global warping, we propose a "local" warping method which includes the following two rules: (a) Each organ on the target person's face (e.g., eye and mouth) is warped independently without considering its relationship with the rest of the face. (b) The feature positions for each organ are warped in a way to make the organ's absolute shape as close as possible to the organ in the source person's expression face. The easiest way to perform "local" warping is to copy the feature positions of one particular organ from the source person's face to the target person's face. By this way, the organ shapes in the target person's face can be exactly the same as the ones on the source person's face. However, since different people have quite different appearances in their faces, it is difficult to decide the proper location, size, and orientation of the "copied" organ. Therefore, in this paper, we propose to use a simple but effective method to perform local warping. That is, we locally "re-shape" the feature points for each organ from the global warping results to make the re-shaped organ similar to the one in the source person's expression face. In this paper, we use the organ's height-to-width ratio as the constraints to perform local re-shaping. For example, Figs 3 and 4 show the local "re-shaping" process of the feature positions for the mouth. In Figs 3 and 4, (a) is the mouth feature positions for the source person's expression face and (b) is the mouth feature positions for the target person's global warped face. Our target is to "re-shape" the global warped mouth feature positions in (b) such that the mouth's height-to-width ratio can be the same as the source person's mouth in (a).

Fig. 3 shows the height re-shaping process. In Fig. 3, we first identify the horizontal middle line of the bounding box surrounding the mouth feature positions (i.e., the horizontal dashed red line in Fig. 3 (a) and (b)). After that, the feature positions in (b) will be vertically moved with respect to this middle line. For example, the feature position $P_t$ in Fig. 3 (b) is vertically moved by Eqn. (1).

$$d_{t1} = d_{s1} \cdot \frac{W_t}{W_s} \tag{1}$$

where $d_{t1}$ is the vertical distance between $P_t$ and the middle line in Fig. 3 (b). $d_{s1}$ is the vertical distance between $P_s$ and the middle line in Fig. 3 (a) where $P_s$ is $P_t$'s corresponding feature position in the source person's mouth. And $W_s$ and $W_t$ are the width of the mouth's minimum bounding box in Fig. 3 (a) and (b), respectively.

Similarly, Fig. 4 shows the width re-shaping process of the target person's mouth where the horizontal locations of the feature position $P_t$ can be moved by:

$$d_{t2} = d_{s2} \cdot \frac{H_t}{H_s} \qquad (2)$$

where $d_{t2}$ is the horizontal distance between $P_t$ and the vertical middle line of the minimum bounding box (i.e., the red dashed line in Fig. 4 (b)). $d_{s2}$ is the horizontal distance between $P_s$ and the vertical middle line in Fig. 4 (a). $H_s$ and $H_t$ are height of the mouth's minimum bounding box in Fig. IV (a) and (b), respectively.

By Eqns (1) and (2), we can move all the feature positions on the target person's mouth and reshape its height-to-width ratio to be the same as the source person's expression mouth. Furthermore, the same process can be applied to re-shape the other organs such as the eyes, the eyebrows, and the nose. By this way, the shape of each organ can be adjusted close to the one in the source person's expression face.

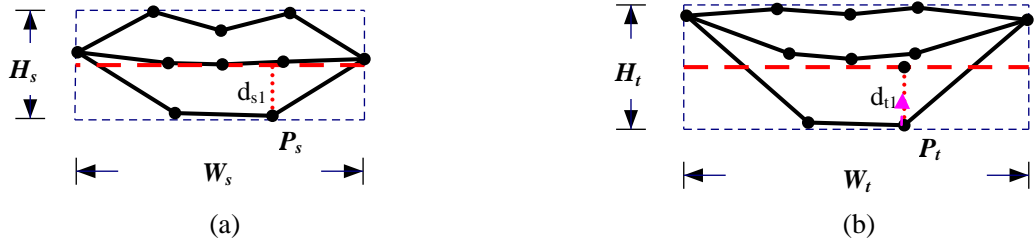

(a)  (b)

Fig. 3 Local re-shaping process of the mouth in vertical direction (the feature point $P_t$ is moved vertically to make the height-to-width ratio of (b) close to (a)). (a) The mouth feature positions of the source person's expression face, (b) The mouth feature positions of the globally warped target person expression face (the blue dashed box is the minimum bounding box surrounding the mouth feature positions and the red dashed line is the horizontal middle line of the minimum bounding box).

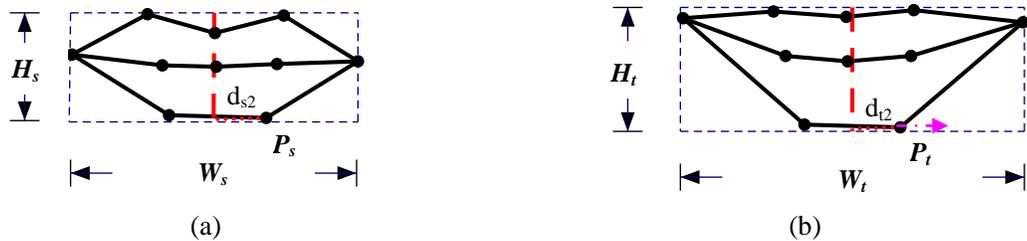

(a)  (b)

Fig. 4 Local re-shaping process of the mouth in horizontal direction (the feature point $P_t$ is moved horizontally to make the height-to-width ratio of (b) close to (a)). (a) The mouth feature positions of the source person's expression face, (b) The mouth feature positions of the globally warped target person expression face (the blue dashed box is the minimum bounding box surrounding the mouth feature positions and the red dashed line is the vertical middle line of the minimum bounding box).

Furthermore, two things need to be noted about local warping: (a) When the feature positions of one organ are warped, the other feature positions in the target person's face remain unchanged. Therefore, each organ can be geometrically warped locally and independently. (b) Since the relationship between the feature points and their corresponding organs in the face can be automatically decided by many automatic feature extraction methods [3,12], this local warping step as well as the entire process of our proposed expression cloning algorithm can be performed automatically in practice.

Compared with global warping which tries to make the relative neutral-to-expression feature position movements to be the same [1-6,8], the local warping tries to equalize the "absolute" organ shapes between the source and the target expression faces. By this way, even if the organ appearances are different between people, the organs of the target person's expression face can still be warped similar to the ones in the source person. Fig. 2 (d) is the result by our proposed local warping method from the three available images in Fig. 1. Compared with Fig. 2 (c), organs such as the mouth and the eyes in (d) are more similar to those of the source person in Fig. 2 (a), thus making the smile expression more recognizable. However, since the local warping method does not consider the relationship among different organs and the neutral-to-expression ratio among organs, the locally warped results still look unnatural. For example, the mouth in (d) is too small and incoherent with the other parts in the target person's face. Therefore, new methods need to be proposed to combine both the local and global warping results. In this paper, we propose a new elastic model to balance the effects of global and local warping. The proposed elastic model can not only create more vivid results for each local organ, but also keep the reasonable global relationship among organs.

Furthermore, besides geometric warping, the detailed illumination information is also important in creating good expression results. Liu et al. [5] propose expression ratio image (ERI) which transfers the illumination detail information based on the relative pixel-value ratio between the neutral and the expression face images. Although this simple method can create effective results, it has the following limitations: (a) Since the facial features are different for different people, simply applying the relative pixel-value ratio from one face to another may not be able to transfer details efficiently. (b) Since ERI may include noise, the transferred illumination details have to be filtered. However, since the filtering in ERI does not consider the structural characters of the face, significant illumination details may be weakened after filtering. Although Song et al. [2] improve ERI by mapping all details into a mesh image to reduce the facial feature differences, the facial structure characters are still not considered when evaluating illumination details, thus leading to unsatisfactory results in some cases. Therefore, in this paper, we further propose a muscle-distribution-based (MD) model which utilizes the muscle distribution information to evaluate the importance of the facial illumination details. Since the MD model includes the muscle distribution to reflect the facial structure, the illumination details transferred by our MD model are more precise than the previous methods in [2,5].

Based on the above discussions, we can propose a new elastic-plus-muscle-distribution-based (E+MD) algorithm which uses the elastic model for geometric warping and the MD model for transferring illumination details. The proposed E+MD algorithm is described in detail in the following.

III. THE ELASTIC-PLUS-MUSCLE-DISTRIBUTION-BASED ALGORITHM

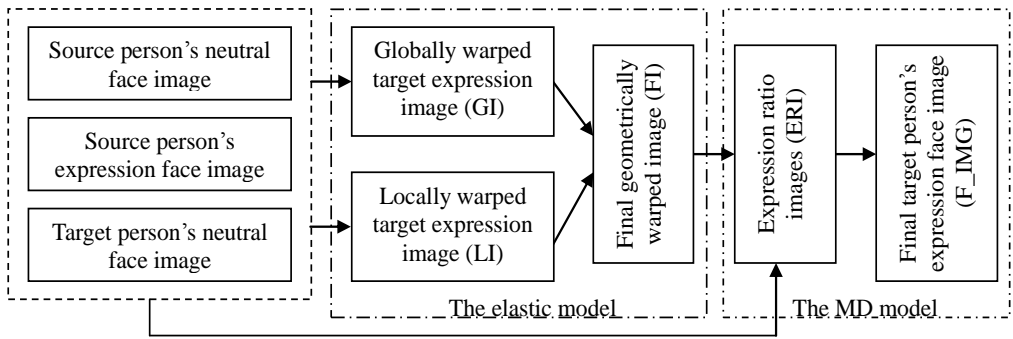

Fig. 5 The framework of the E+MD algorithm.

The framework of our proposed E+MD algorithm is shown in Fig. 5. In Fig. 5, the three available images (i.e., the source person's neutral face, source person's expression face, and target person's neutral face) are first used to create a globally warped target expression image (GI) [3,5] and a locally warped target expression image (LI). Then, our proposed elastic model is used to balance the effects of GI and LI for creating the final geometrically warped image (FI). After FI is achieved, it is aligned with the three available images to create the initial ERIs [5]. These ERIs will be further processed by our MD model for achieving the final target person's expression face image (F_IMG) which will include the facial illumination details. It should be noted that: (1) Although the warping method in [3,5] is used to create the global warping result in this paper, our elastic model is general and it can be combined with the other global warping methods [8,15] to achieve the improved facial expression results. This point will be further discussed in the experimental result section. (2) As discussed in the previous sections, the elastic model and the MD model are the key components in our E+MD algorithm. Thus, in the following, we will describe these two models in detail.

*A. The elastic model*

As mentioned above, the target for our elastic model is to balance the effects of both the global warping result GI (e.g., Fig. 2 (c)) and the local warping result LI (e.g., Fig. 2 (d)), thus to give a more natural and vivid expression in the target person's face.

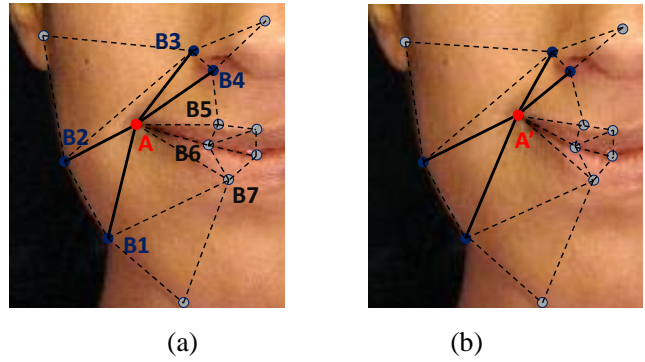

(a)                  (b)

Fig. 6 The feature position of the mouth corner and its neighboring feature points (*B1* to *B4*). (a) *A* is the location of global warping result, (b) *A'* is the location of local warping result. (Best viewed in color)

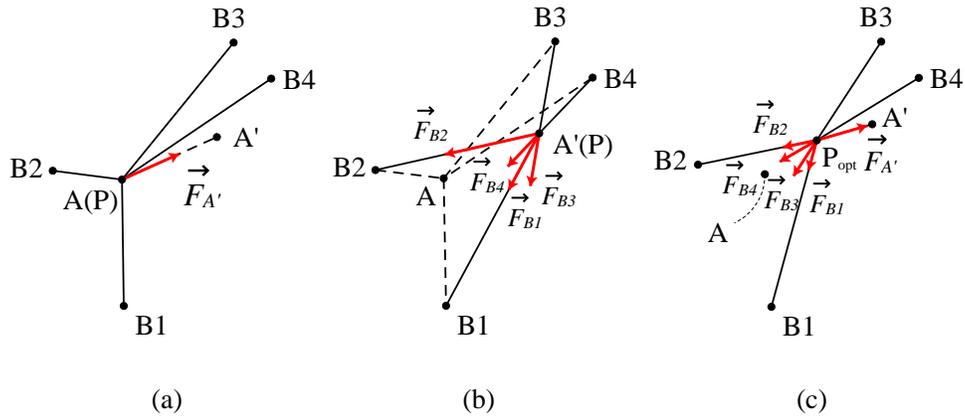

(a)               (b)              (c)

Fig. 7 Elastic forces for different *P* locations (the red arrows are the spring forces to *P*). (a) *P* is at the global warping result location *A*, (b) *P* is at the local warping result location *A'*, (c) *P* is at the final balanced location $P_{opt}$.

Our proposed elastic model can be described by Fig. 6 and Fig. 7. In Fig. 6, *A* is the location of the mouth corner in the target person's expression face calculated by global warping (i.e., the location of *P* in GI) and *A'* is the location of the mouth corner decided by local warping (i.e., the location of *P* in LI). *B1-B7* are the neighboring feature positions of *A* in the global warping result. These neighboring relationships are decided by triangulation as in Fig. 2 (b). And our proposed elastic model will create an optimized geometric position of the mouth corner based on *B1-B7* as well as *A* and *A'*. In order to ease the discussion, in the following description, we will omit the effects of *B5-B7* and focus on discussing how to achieve the optimal mouth corner location only by its four neighboring points (i.e., *B1-B4*). Note that the omission of *B5-B7* is only to simplify the description in Fig. 7. In our real processing, all the neighboring points *B1-B7* are used to balance the location of *P*.

For the convenience to show the process of using our elastic model, we redraw the points *B1-B4* and *A, A'* as well as the lines linking them in Fig. 7. Also, we use *P* to indicate the current location of the mouth corner point in Fig. 7.

We assume that there is a spring connected between the mouth corner point *P* and each of its neighboring feature positions (i.e., *B1-B4*), respectively. At the same time, there is also another spring connected between the mouth corner point *P* and its local position *A'*. The springs can generate elastic forces. The neighboring feature positions (i.e, *B1-B4*) try to pull or push the mouth corner point to its global warping position *A* while the force between the mouth corner point and *A'* tries to pull or push *P* to its local warping position *A'*. Since the positions of *A* and *B1-B4* are all calculated by global warping, there will be no elastic forces from *B1-B4* when the mouth corner point is located in its global warping location *A* (i.e., when *P* is overlapping with *A*, as in Fig. 7 (a)). However, the force from *A'* will pull the mouth corner point from *A* to *A'* (as shown by the red arrows in Fig. 7 (a)). On the other side, when the mouth corner point is located in its local warping location *A'* (Fig. 7 (b)), there will be no elastic force from *A'* as *P* is overlapping with *A'*, but the forces from *B1-B4* will try to pull or push *P* back to its global warping location *A*. Finally, the optimal location of the mouth corner point $P_{opt}$ (i.e., the location of *P* in FI) can be decided where the elastic forces from different directions are balanced, as in Fig. 7 (c) and Eqn. (3):

$$P = P_{opt} \quad when \quad \sum_i \vec{F}_{Bi}(P) + \vec{F}_{A'}(P) = 0 \qquad (3)$$

where $\vec{F}_{Bi}(P)$ and $\vec{F}_{A'}(P)$ are the elastic forces to the current position *P* from the neighboring feature position $B_i$ and the local warping location *A'*, respectively. The elastic forces $\vec{F}_{Bi}(P)$ and $\vec{F}_{A'}(P)$ can be calculated by Hooke's law [13], which tells the force generated by a spring is in direct proportion with the extension of that spring, as in Eqn. (4):

$$\begin{cases} \vec{F}_{Bi}(P) = k_{Bi} \cdot \Delta l_{Bi}(P) \cdot \vec{e}_{Bi}(P) \\ \vec{F}_{A'}(P) = k_{A'} \cdot \Delta l_{A'}(P) \cdot \vec{e}_{A'}(P) \end{cases} \quad (4)$$

where $\Delta l_X(P)$ (*X=A* or *Bi*) denotes the distance change between *P* and *X* when *P* moves. $\Delta l_{Bi}(P) = 0$ when *P* is in the position of *A* and $\Delta l_{A'}(P) = 0$ when *P* is in the position of *A'*. $\vec{e}_X(P)$ is the unit vector pointing from *P* to *X* and $k_X$ is the elasticity coefficient for the elastic force between *P* and *X*. In this paper, we let the neighboring points have the same coefficient value (i.e., $k_{B1} = k_{B2} = k_{B3} = k_{B4} = k_N$).

From Eqn. (3), (4) and Fig. 6 and 7, we can see that our elastic model introduces elastic forces to model the effects of the local and global warping results (i.e., GI and LI). Thus, by balancing the impacts from these forces, the advantages of the global and local warping effects can be effectively combined.

In order to speed up the process and avoid large facial cloning distortions, we assume that the optimal position $P_{opt}$ should not be far from its global location *A* and the local location *A'*. Therefore, we can define a searching window *Win* which is bounded by the global location *A* and the local location *A'* (i.e., the grey box in Fig. 8) and only search for $P_{opt}$ within this searching window. Thus, Eqn. (3) can be solved by:

$$P_{opt} = \arg \min_{P, P \in Win} \left\| \sum_i \vec{F}_{Bi}(P) + \vec{F}_{A'}(P) \right\| \quad (5)$$

where *Win* is the searching window bounded by the global location *A* and the local location *A'*, and $\| \cdot \|$ is the Euclidean distance operation. Fig. 8 shows the detailed process for searching for $P_{opt}$. It should be noted that since the distance between *A* and *A'* is normally small, we only need to search for a small window for each feature position. Thus, the computing cost for this step is small.

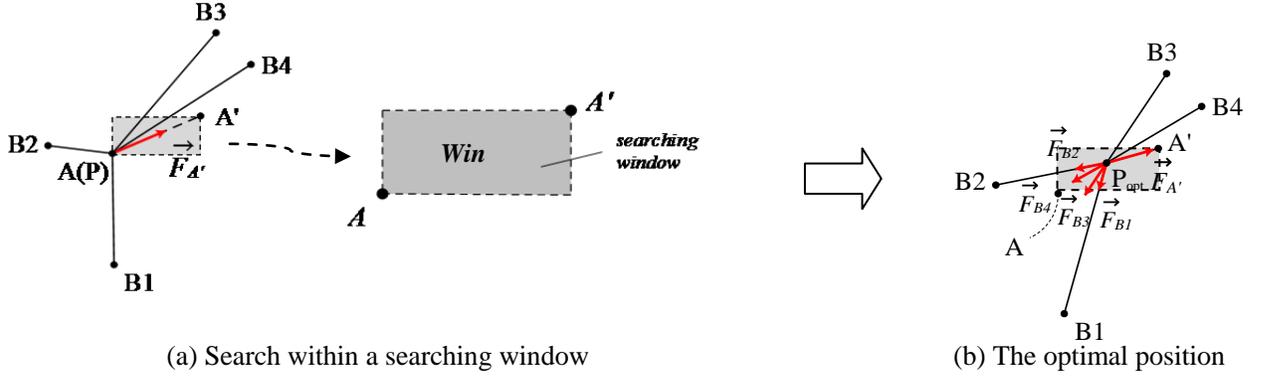

(a) Search within a searching window      (b) The optimal position

Fig. 8 Searching in a searching window bounded by the global location *A* and the local location *A'*.

Furthermore, from Eqn. (4), we can see that the elasticity coefficients $k_X$ is another key issue to decide the relative importance between the global and local warping results. Different ratios between the elasticity coefficients $k_{A'}$ and $k_N$ will lead to different balanced results. In this paper, we define the elasticity ratio as $\lambda_k$, as in Eqn. (6):

$$\lambda_k = \frac{k_{A'}}{k_N} \tag{6}$$

By changing the value of $\lambda_k$, we can control the relative importance between the global and local warped results and create different facial expression cloning results. For example, if we choose a large $k_{A'}$ and a small $k_N$, that is, $\lambda_k$ is large, the final balanced result will be more favorable to the local warping results (e.g., the result will be exactly the same as the local warping result when $k_N= 0$). Therefore, it is very important to select a suitable set of $k_{A'}$ and $k_N$ for achieving optimized result. In this paper, we propose a new distance-based (DB) metric for automatically select the best $\lambda_k$. It is described in the following subsection.

*B. The distance-based metric*

The proposed DB metric for the synthesized target person's expression face image $I_{target\_exp}$ can be described by Eqn. (7):

$$DB(I_{target\_exp}(\lambda_k)) = e^{\|\vec{E}_{target\_exp}(\lambda_k) - \vec{E}_{source\_exp}\|} + \omega_{db} \cdot e^{\|\vec{E}_{target\_exp}(\lambda_k) - \vec{E}_{target\_gl}\|} \tag{7}$$

where $\vec{E}_{target\_exp}(\lambda_k)$ is the feature vector for the synthesized target person's expression face image $I_{target\_exp}(\lambda_k)$ (note that both $I_{target\_exp}$ and $\vec{E}_{target\_exp}$ are controlled by the elasticity ratio $\lambda_k$ in Eqn. (6)), $\vec{E}_{source\_exp}$ is the feature vector for the source person's expression face image, and $\vec{E}_{target\_gl}$ is the feature vector for the target person's global warped expression image. $\|\cdot\|$ is the Euclidean distance operation. And $\omega_{db}$ is the balancing parameter and it is set to be 1 in our experiments.

Since the target for facial expression cloning is to transfer the source person's expression to the target person, better expression cloning result should be expected if the target person's expression is more similar to that of the source person. Therefore, the first term of Eqn. (7) is used to calculate the distance between the target person's final expression face and the source person's expression face. It is straightforward that better expression cloning results should achieve smaller value in the first term.

However, the first term only considers the similarity between the target expression and the source expression while the target person's own characters are not considered. This will lead to a problem that any expression face similar to the source person's expression but totally different from the target person's face will achieve small values in the first term (e.g., if we view the source person's expression face as the final synthesized result, the first term in Eqn. (7) will be 0). Therefore, in our DB metric, we also introduce the second term to calculate the distance between the final target person's expression face and the target person's global warping result (i.e., $\|\vec{E}_{target\_exp} - \vec{E}_{target\_gl}\|$). By taking the target person's global warping result as the reference, the target person's own expression character can be suitably included. Therefore, according to Eqn. (7), an expression cloning result $I_{target\_exp}$ is good if: (a) it can efficiently carry the source person's expression (i.e., close to the source person's expression face), and (b) it suitably holds the target person's own character (i.e., close to the target person's global warping results).

There can be many ways to extract the feature vectors for describing faces in Eqn. (7) (i.e., $\vec{E}_{target\_exp}$, $\vec{E}_{source\_exp}$, and $\vec{E}_{target\_gl}$). In this paper, we use the Eigenface method [11, 19] for extracting feature vectors. That is, we set images with different expressions of one person as the training set (12 images in our experiments) and perform Principle Component Analysis (PCA) to extract the principal components. Then, for the input images, their eigenvectors will be extracted based on these principal components and used as the feature vectors [11, 19].

With the DB metric in Eqn. (7), the optimal elasticity coefficients $k_{A'}$ and $k_N$ in the elastic model in Eqn. (4) can be decided by:

$$\lambda_{k\_opt} = \arg \min_{\lambda_k} DB(I_{target\_exp}(\lambda_k)) \tag{8}$$

where $\lambda_k = k_{A'}/k_N$ is the elasticity ratio as defined in Eqn. (6). $I_{target\_exp}(\lambda_k)$ is the resulting expression cloning image with elasticity ratio $\lambda_k$. After the optimal elasticity ratio is decided, the optimal elasticity coefficients can be achieved accordingly.

*C. The Muscle-distribution-based (MD) model*

The target for the MD model is to process the detailed illumination information (i.e., ERI in Fig. 1) such that suitable illumination details can be added to the target person's expression face image F_IMG. We observe that facial muscles play key roles in a person's expression details. According to the muscle theory [1,10], most facial illumination details come from the movements of muscles in one's face (e.g., shrink and expand). Therefore, in our MD model, we introduce the muscle distribution information to evaluate the importance of illumination details such that important illumination details are strengthened properly to create more vivid expression results.

Based on the muscle structure theory [10] (as in Fig. 9 (a)), we first define the key muscle areas which are closely related with the illumination details and the muscle movements on person's face. Fig. 9 (b) shows some examples of the defined key muscle areas. Note that in order to show the key muscle areas clearly, only parts of the areas are displayed in Fig. 9 (b). The locations of the key muscle areas can be decided by one feature point (e.g., the muscle area *B* in Fig. 10 (a)) or the center of several feature points (e.g., the muscle area *A* in Fig. 10 (a)). Furthermore, the radius of the key muscle areas can also be decided by the distance among the feature positions. As mentioned, since the feature positions in the face can be automatically decided [3, 12], these key muscle areas can be automatically calculated from the feature positions as long as the corresponding relation between the muscle areas and feature positions has been preset.

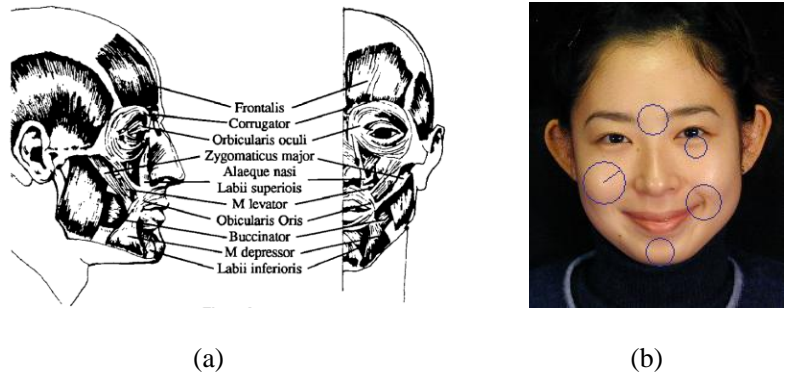

(a)                 (b)

Fig. 9 (a): The muscle structure [1,10]. (b): Some example key muscle areas.

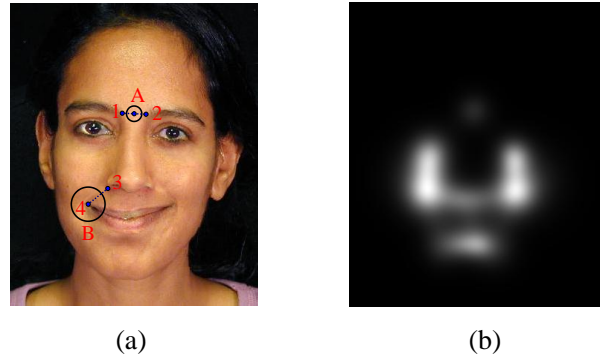

(a)                 (b)

Fig. 10 (a) Deciding the key muscle areas by feature points, (b) An example mask image for an entire face image (the brighter pixels indicate more illumination details are enhanced).

After the key muscle areas are decided, our MD model will strengthen the illumination details inside these key muscle areas. The final processed illumination detail at pixel ($u$, $v$) by our MD model can be calculated by Eqn. (9).

$$IL^{MD}(u,v) = M(u,v) IL(u,v) \qquad (9)$$

where $IL(u, v) = ERI(u, v) - 1$ is the shifted ERI [5] and $M(u, v)$ is the detail importance mask decided by the key muscle areas. Since the main purpose of our MD model is to strengthen the illumination details inside the key muscle areas, the detail importance mask $M(u, v)$ should mainly obey the following three rules: (a) The illumination details should be strengthened properly inside these key muscle areas. (b) The illumination details outside these circles should not be too much affected. (c) The changes of illumination details should be as smooth as possible at the boundary of these circles and no abrupt changes are allowed. Considering the above three rules, we use a low-pass-filter-like mask for each key muscle area and let it fade gradually around the boundary of the area, as in Eqn. (10).

$$M(u,v) = 1 + h(u_0, v_0) \cdot e^{-\frac{(u-u_0)^2 + (v-v_0)^2}{2\sigma(u_0,v_0)^2}} \tag{10}$$

where $(u_0, v_0)$ is the center of the key muscle areas. $h(u_0, v_0)$ determines the strength by which illumination details are added and it is adaptive for different key muscle areas. Furthermore, $\sigma(u_0, v_0)$ is the impact range of the key muscle area centered at $(u_0, v_0)$. In this paper, $\sigma(u_0, v_0)$ is calculated by:

$$\sigma(u_0, v_0) = \frac{1}{\sqrt{ln\,2}} r(u_0, v_0) \tag{11}$$

where $r(u_0, v_0)$ is the radius of the area as in Fig. 10 (a). Fig. 10 (b) shows an example of the mask image for an entire face where the brighter pixels indicate that more illumination details are enhanced.

By our proposed elastic model and MD model, the facial expression cloning results can be obviously improved. The experimental results are shown in the next section.

## IV. EXPERIMENTAL RESULTS

In this section, we show experimental results for our proposed E+MD algorithm.

Fig. 11 shows the results of the experiment in Fig. 1 where the following seven methods are compared: (a) use only global warping for facial expression cloning (Global warping) [3], (b) use only local warping for facial expression cloning (Local warping), (c) use only our elastic model for balanced result, (d) use global warping for geometric warping and use ERI for adding illumination details (Global warping+ERI) [5], (e) use local warping for geometric warping and ERI for the expression cloning (Local warping+ERI), (f) use our elastic model for geometric warping and ERI for adding illumination details (Elastic model+ERI), and (g) use our elastic model for geometric warping and use our MD model for adding illumination details (E+MD). In order for easy comparison, we also attach the original person's expression face in Fig. 11 (h).

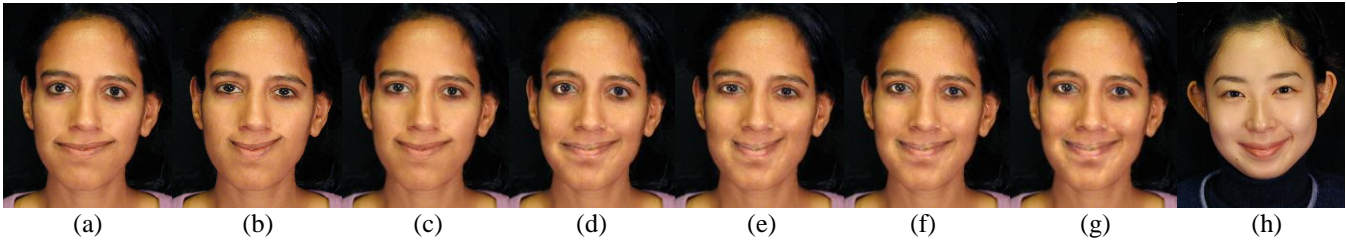

(a)　　　(b)　　　(c)　　　(d)　　　(e)　　　(f)　　　(g)　　　(h)

Fig. 11 Facial Expression Cloning Results for the Experiment in Fig. 1. ((a): Global warping [3], (b) Local warping, (c) Elastic model, (d) Global warping+ERI [5], (e) Local warping+ERI, (f) Elastic model +ERI, (g) E+MD algorithm, (h) the source person's expression face) (best viewed in color)

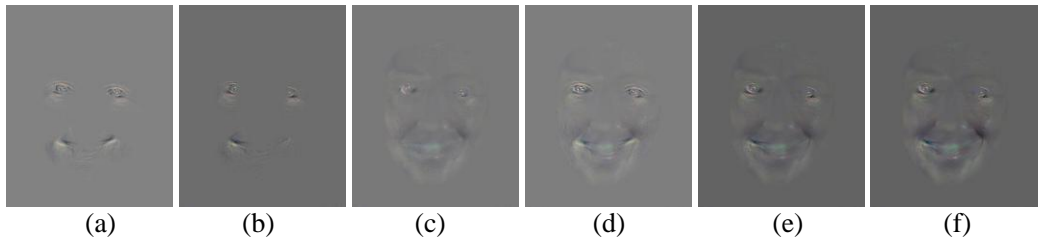

(a)　　　(b)　　　(c)　　　(d)　　　(e)　　　(f)

Fig. 12 The difference images of the expressions in Fig. 11. ((a)~(f) correspond to the differences between Figs 11 (b)~(g) to Fig. 11 (a))

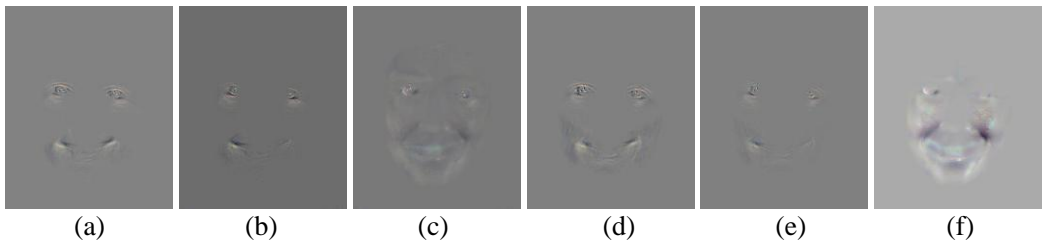

(a)　　　(b)　　　(c)　　　(d)　　　(e)　　　(f)

Fig. 13 The difference images of the expressions in Fig. 11. (**(a)**: Fig. 11 (b) – Fig. 11 (a), **(b)**: Fig. 11 (c) – Fig. 11 (a), **(c)**: Fig. 11 (d) – Fig. 11 (a), **(d)**: Fig. 11 (e) – Fig. 11 (d), **(e)**: Fig. 11 (f) – Fig. 11 (d), **(f)**: Fig. 11 (g) – Fig. 11 (f))

From Fig. 11, we can see that the results by only using the global or local warping (i.e., Figs 11 (a) and (b)) are obviously less similar to the "smile" expression of the source person. Although the "global warping+ERI" and "local warping+ERI" methods improve the result by adding illumination details (i.e., Figs 11 (d) and (e)), their expressions are still less satisfactory as the smile extent is obviously different from that of the source person in Fig. 1. The smile extent is more natural in Fig. 11 (f) where the local facial difference between the source and the target people are reduced by our elastic model. Finally, by adding illumination details using our MD model, it is obvious that the smile expression in Fig. 11 (g) is the most expressive and the most effective in cloning the source person's facial expression.

In order to further analyze the results, we also create the difference images among the expression images in Fig. 11. The difference images are created by taking the difference between two images and normalizing the values to the range 0~255. The resulting difference images are shown in Figs 12 and 13. Fig. 12 shows the difference images by taking the difference between Figs 11 (b)~(g) and the global warping result in Fig. 11 (a), and Fig. 13 shows the difference images between different image pairs in Fig. 11. From Figs 12 (b) and (f), we can see that our E+MD algorithm improves the expression results by suitably including geometric warping adjustments and illumination details. Also, from Fig. 13 (f), it is clear that illumination details of the key muscle areas are properly enhanced by our MD model.

Furthermore, Figs 14-15 show the other two sets of facial expression cloning results. In Figs 14-15, we want to transfer one person's expression in (e) to the person in (a). Figs 14-15 (b), (c), and (d) show the results of the "global warping+ERI" method [5], use our elastic model for geometric warping and ERI for adding illumination details, and our E+MD algorithm, respectively. It is obvious that the result by our E+MD algorithm catches the sad expression more precisely where the organs such as the eyes and the mouth in (d) are more vivid than those in (b). Furthermore, comparing (d) and (c), we can see that our MD model can effectively strengthen the details (e.g., the details between the eyebrows and around the nose) to make the expression more recognizable.

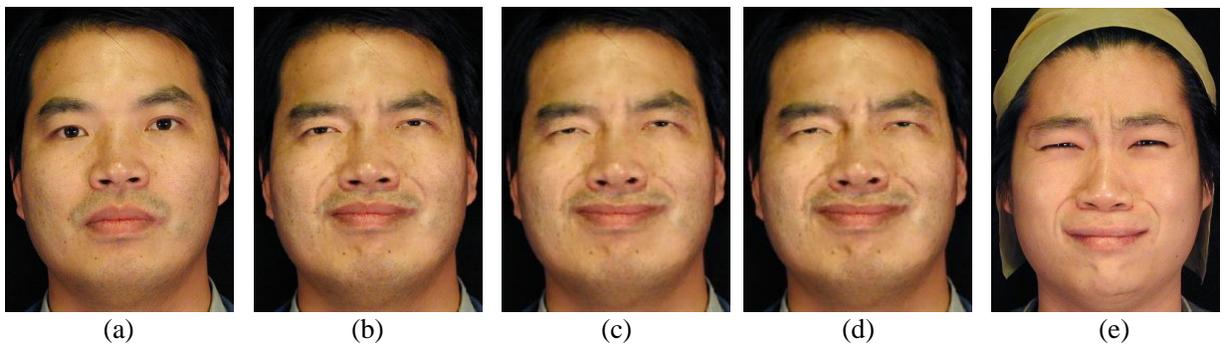

(a) (b) (c) (d) (e)

Fig. 14 Facial expression cloning results. ((a): Target person's neutral face, (b) Global warping+ERI [5], (c) Elastic model+ERI, (d) Our E+MD algorithm, (e) Source person's expression face) (best viewed in color)

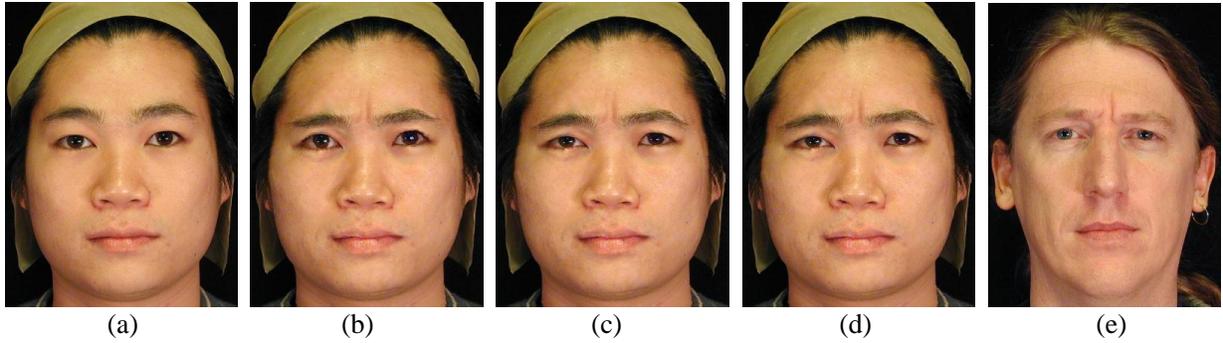

Fig. 15 Facial expression cloning results. ((a): Target person's neutral face, (b) Global warping+ERI [5], (c) Elastic model+ERI, (d) Our E+MD algorithm, (e) Source person's expression face) (best viewed in color)

In order to further demonstrate the effectiveness of our algorithm, we also compare our results with Song's method [2] where the vertex tent coordinate (VTC) transfer and mesh images are used. In Fig. 16 (f) and Fig. 17 (d), the expressions of the source persons are to be transferred. Fig. 16 (c) and Fig. 17 (b) are the expression cloning results of Song's method [2] while Fig. 16 (e) and Fig. 17 (d) are the results of our E+MD algorithm. Again, it is obvious that our E+MD algorithm creates the best result. Comparing the results in Figs 16-17, we can also see that our E+MD algorithm can produce more satisfactory results than Song's method [2] (e.g., the shape of the mouth and the details around the eyebrow center in Fig. 16 (e) are more precise than (c)). This is because: (a) although Song's method [2] utilizes 3D model and mesh image to reduce the facial differences, it is still less efficient in addressing the impacts from local facial feature differences. (b) Song's method [2] also ignores the facial structure information during illumination detail transfer. Compared with [2], these problems can be suitably addressed by our proposed elastic and MD models.

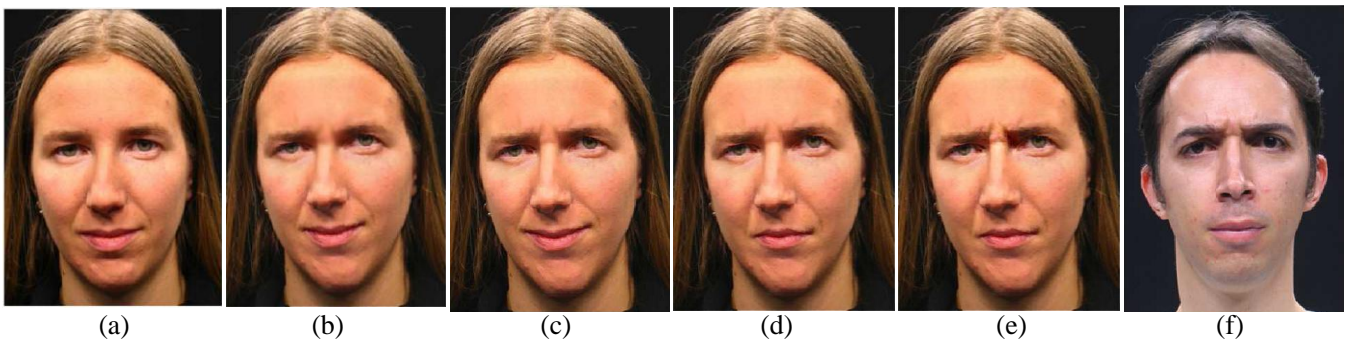

Fig. 16 Facial expression cloning results. ((a): Target person's neutral face, (b) Global warping+ERI [5], (c) Song's method [2], (d) Elastic+ERI, (e) Our E+MD algorithm, (f) Source person's expression face) (best viewed in color)

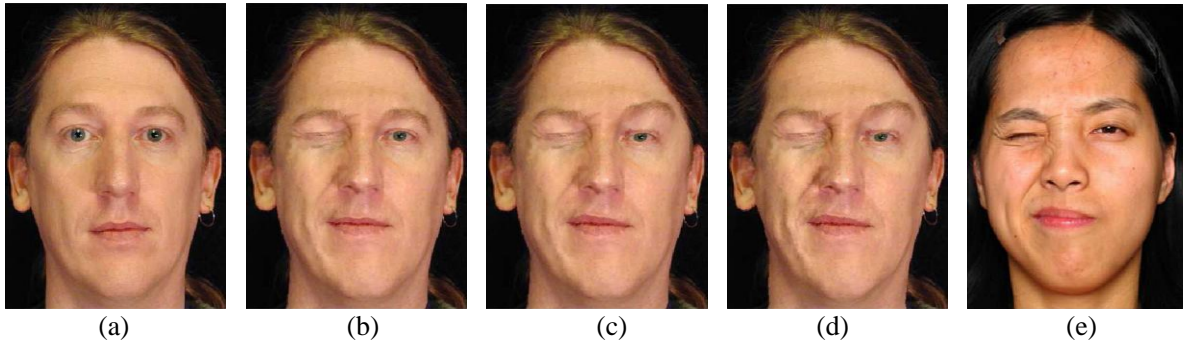

(a)           (b)           (c)           (d)           (e)

Fig. 17 Facial expression cloning results. ((a): Target person's neutral face, (b) Song's method [2], (c) Elasic+ERI by using Song's method for global warping, (d) our E+MD algorithm by using Song's method for global warping, (e) Source person's expression face) (best viewed in color)

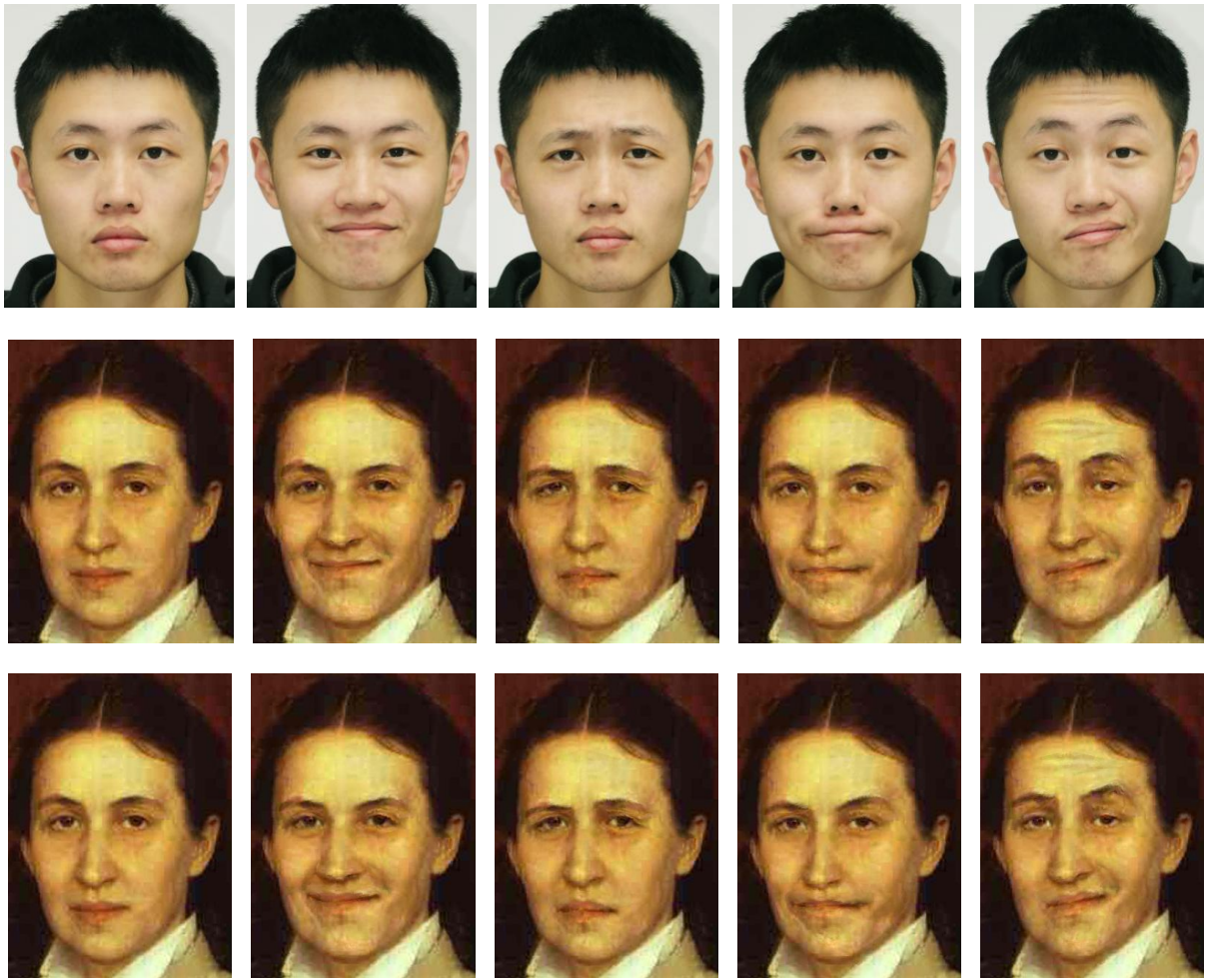

Fig. 18 Using our E+MD algorithm to transfer the expressions of a person in a video to the neutral face of a person in portrait (First row: the source person's expression video; Second row: the target person's expression video whose feature positions are achieved manually; Third row: the target person's expression video whose feature positions are achieved by the automatic AAM method [12]).

Furthermore, Fig. 18 shows another example by using our E+MD algorithm to transfer the expressions of a person in a video to a neutral face of a portrait. In Fig. 18, the first row is the source person's expression sequence. The second row is the target person's expression sequence whose feature positions are achieved by manually identifying the feature positions. And the third row is the target person's expression sequence whose feature positions are achieved by an automatic feature position identification method (i.e., the Active Appearance Model (AAM) method [12]). From Fig. 18, we can see that the expressions of the source person are efficiently transferred to the person in the portrait. Besides, the results in the third row are similar to the ones in the second row, this further demonstrates that the automatic feature position identification methods [12, 15] can create similar results as the manual feature position identification method. Note that the framework of our algorithm is general, and besides the AAM method [12], our algorithm can also be combined with other advanced automatic feature position identification methods [25-26] to perform automatic facial expression cloning in more challenging conditions (e.g., poor lighting condition or low image resolution).

Moreover, as mentioned, in the experiments of this paper, our E+MD algorithm is implemented based on the geometric warping strategy [3,5]. However, the idea of our proposed algorithm is general and it can be extended to combine with other facial expression warping strategies [3-4,8,15-16,20]. For example, we can use the Radial-Basis-Function-based method [15] or the deformation-transfer-based method [8] to create the globally warped results and then apply our E+MD algorithm accordingly. Figs 19 and 20 show two examples by applying our elastic model on the warping results of [15]. In Figs 19-20, (a) is the target person's neutral face and (d) is the source person's expression face. (b) is the result by [15] and (c) is the result by applying our elastic model on (b). Note that our MD model is not applied in Figs 19-20 since the target person's face does not include the texture information. From Figs 19-20, it is clear that our elastic model can effectively improve the facial expression results of [15] by introducing local warping to handle the local organ similarities (e.g., compared with (b), the mouth and the eyebrow regions in (c) are more vivid and closer to those of the source person's expression face in (d)).

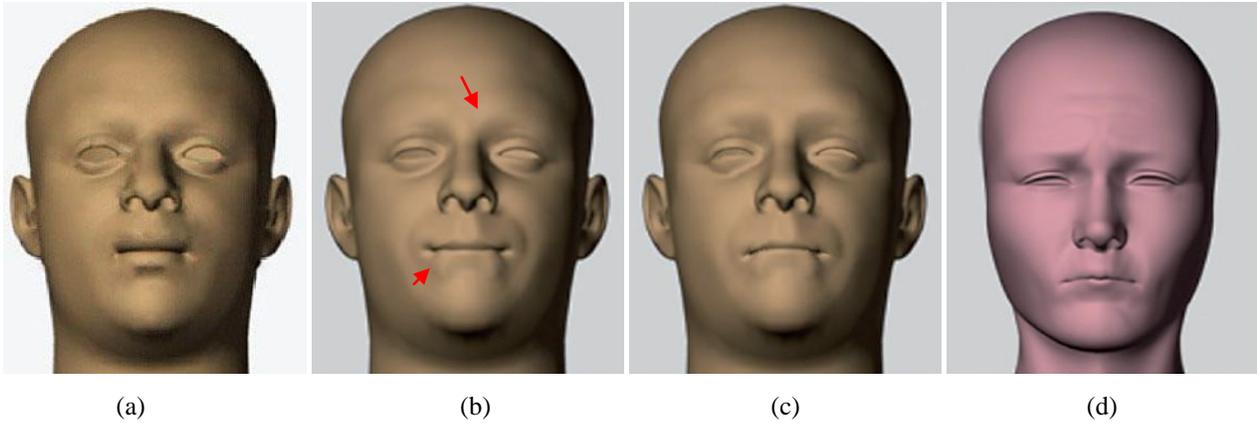

(a)　　　　　　　　　(b)　　　　　　　　　(c)　　　　　　　　　(d)

Fig. 19 Facial expression cloning results by applying our algorithm on other expression warping methods. ((a) Target person's neutral face, (b) Result by Seol's method [15], (c) Result by applying our elastic model on (b), (d) Source person's expression face) (best viewed in color)

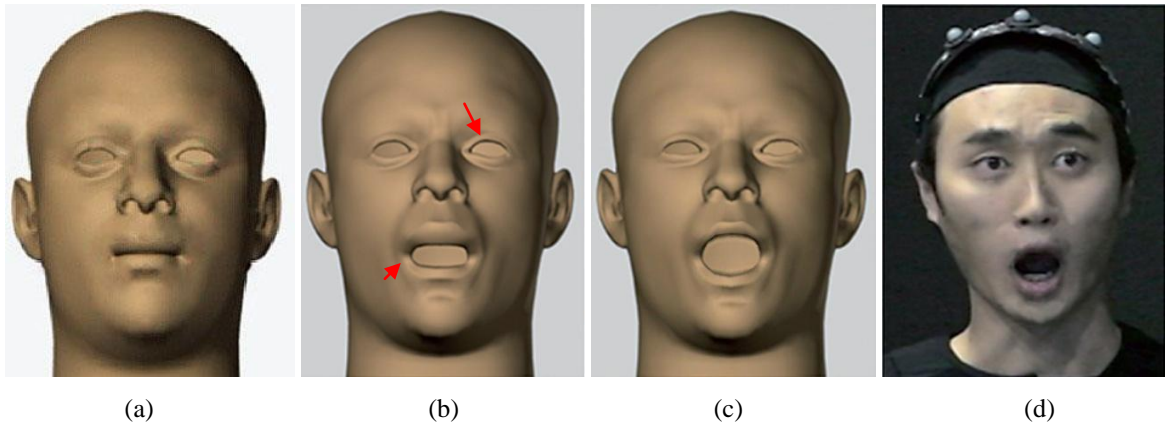

(a)　　　　　　　　　(b)　　　　　　　　　(c)　　　　　　　　　(d)

Fig. 20 Facial expression cloning results by applying our algorithm on other expression warping methods. ((a) Target person's neutral face, (b) Result by Seol's method [15], (c) Result by applying our elastic model on (b), (d) Source person's expression face) (best viewed in color)

Furthermore, Fig. 21 show another example by combining our elastic model with the method in [20] (i.e., using the facial transfer results in [20] as the global warping results and then applying our elastic model to obtain better expression transfer results). Figs 21 further demonstrates that our proposed model can effectively provide better expression transfer results when combining with the existing facial expression cloning methods (e.g., compared with (a), the eye and month regions in (b) are closer to those of the source person's expression face in (c), making the smile expression more vivid).

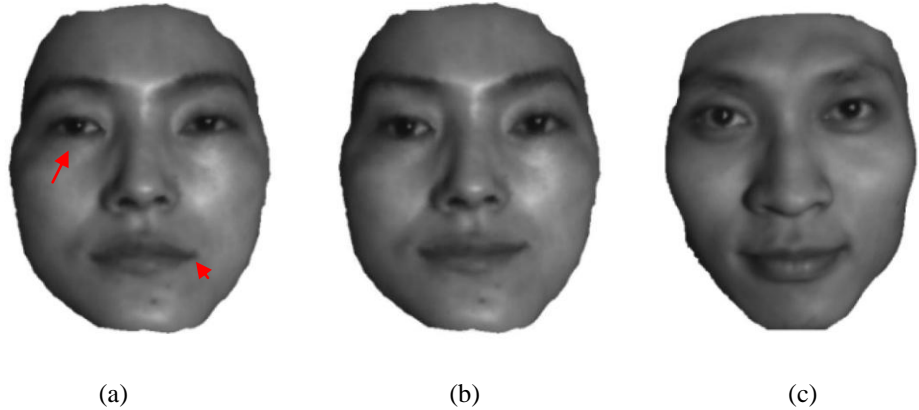

(a) (b) (c)

Fig. 21 Facial expression transfer results by combing our algorithm with the method in [20]. ((a) Target person's expression face by [20], (b) Target person's expression face by combining our algorithm with the method in [20], (c) Source person's expression face).

Table 1 Subjective user study results for different facial expression cloning algorithms

|  | Global warping +ERI [5] | Song's method [2] | Seol's method [15] | Wang's method [20] | Elastic Model+ERI | **E+MD** |
|---|---|---|---|---|---|---|
| Fig. 11 | 2.40 | - | - |  | 3.75 | **3.95** |
| Fig. 14 | 2.50 | - | - |  | 2.85 | **3.00** |
| Fig. 15 | 3.10 | - | - |  | 3.35 | **3.45** |
| Fig. 16 | 2.95 | 2.85 | - |  | 3.85 | **4.10** |
| Fig. 17 | 2.30 | 3.70 | - |  | 3.85 | **3.85** |
| Fig. 19 | - | - | 3.15 |  | - | **3.55** |
| Fig. 20 | - | - | 3.75 |  | - | **3.95** |
| Fig. 21 |  |  |  | 2.90 |  | **4.05** |

Finally, Table 1 shows a subjective user study test on the results in Figs 11- 20 [15, 17-18]. We asked 30 participants to grade the facial expression mapping results for different algorithms. The participants include 18 males and 12 females whose ages ranged from 20-60. All the participants are without visual problems and 28 participants reported that they did not know about facial expression cloning. In order to avoid the evaluation bias, the algorithm information of the resulting images is totally concealed from the participants and the images for different algorithms are randomly placed. During the test, the participants are first required to observe the results. After that, they shall give a score to each facial expression mapping result. The score is within the range of 1~5, where 1 means the poorest and 5 means the best. Finally, the scores over 30 participants are averaged and quantized to be the subjective evaluation score for each for each facial expression cloning image. The user study test result in Table 1 further demonstrates that our proposed E+MD algorithm can improve the facial expression cloning results from the previous methods. Besides, comparing "Elastic Model+ERI" with our E+MD, we can also see that the "E+MD" method has larger values. This demonstrates that adding illumination details can further improve the facial expression cloning results.

## V. CONCLUSION AND FUTURE WORK

In this paper, a new E+MD algorithm for facial expression cloning is proposed. The proposed algorithm introduces an elastic model for balancing the global and local warping effects and a muscle-distribution model for evaluating and strengthening the facial illumination details. In addition, we also propose a new distance-based metric for automatically selecting the parameters to balance the global and local model effects. Experimental results show that our proposed algorithm can achieve better expression results than the existing methods.

The future works for our algorithm will include the following aspects:

(a) Extending to 3D facial expression cloning. In the experiments in this paper, the facial expressions are synthesized with a 2D elastic model. However, our elastic model can be easily extended to the 3D space. For example, we can first achieve the 3D feature positions from the face image or face image sequences [19, 23-24]. Then, the global locations of feature positions in the target person's expression can be obtained. By directly re-shaping these feature positions in the 3D space, we can achieve their 3D local locations. Finally, we can use the elastic model to get the 3D balanced locations for these feature positions and apply 3D warping [21-22] to create the expression cloning results.

(b) Extending to other shape deformation and warping applications. Since the basic idea of our algorithm is to introduce the local model and apply the elastic model to improve the warped feature position locations, it is not limited to facial expression cloning. Instead, it can also be applied to improve the shape deformation results of other objects such as the human or animal body shapes [21-22].

## REFERENCES

[1] K. Kahler, J. Haber, and H. P. Seidel, "Geometry-based muscle modeling for facial animation," *Graphics interface*, pp.37-46, 2001.

[2] M. Song, Z. Dong, C. Theobalt, H. Wang, Z. Liu, H. Seidel, "Generic framework for efficient 2D and 3D facial expression analogy," *IEEE Trans. Multimedia*, vol. 9, no. 7, 1384-1395, 2007.

[3] G. Wolberg, "Digital image warping," *IEEE Comp. Society Press*, 1990.

[4] T. Weise, S. Bouaziz, H. Li, and M. Pauly "Kinect realtime performance-based facial animation," *SIGGRAPH*, pp. 1-8, 2011.

[5] Z. Liu, Y. Shan and Z. Zhang, "Expressive expression mapping with ratio images," *SIGGRAPH*, pp. 271-276, 2001.

[6] B. Theobald, I. Matthews, J. Cohn, and S. Boker, "Real-time expression cloning using active appearance models," *Int'l Conf. Multimodel Interfaces*, pp. 134-139, 2007.

[7] J. Noh and U. Neumann, "Expression cloning," *SIGGRAPH*, pp. 277-288, 2001.

[8] R. W. Sumner and J. Popović, "Deformation transfer for triangle meshes," *ACM Trans. Graphics*, vol. 23, no. 3, pp. 399-405, 2004.

[9] T. Igarashi, T. Moscovich and J. F. Hughes, "As-rigid-as-possible shape manipulation," *ACM Trans. Graphics*, vol. 24, pp. 1134-1141, 2005.

[10] K. Waters, "A muscle model for animating three-dimensional facial expression," *SIGGRAPH*, vol. 21, pp. 17-24, 1987.

[11] M. Turk and A. Pentland, "Eigenfaces for recognition," *Journal of Cognitive Neuroscience*, vol. 3, no. 1. pp. 71-86, 1991.

[12] T. Cootes, G. Edwards and C. Taylor, "Active appearance models," *IEEE Trans. Pattern Anal. Mach. Intell.*, vol. 23, no. 6, pp. 681–685, 2001.

[13] J. Rychlewski, "On Hooke's law," *Journal of Applied Mathematics and Mechanics*, vol. 48, no. 3, pp. 303-314, 1984.

[14] Y. Zhang, W. Lin, B. Sheng, J. Wu, H. Li, C. Zhang "Facial expression mapping based on elastic and muscle-distribution-based models," *IEEE International Symposium on Circuits and Systems (ISCAS)*, pp. 2685-2688, 2012.

[15] Y. Seol and J. Lewis, "Spacetime expression cloning for blendshapes," *ACM Trans. Graphics*, vol. 31, no. 14, pp. 1-12, 2012.

[16] H. Li, T. Weise, M. Pauly, "Example-based facial rigging," *ACM Transactions on Graphics*, vol. 29, no. 4, pp. 1-12, 2010.

[17] M. Sun, Z. Liu, J. Qiu, Z. Zhang, and M. Sinclair, "*Active lighting for video conferencing,*" IEEE Trans. Circuits and Systems for Video Technology, vol.19, no.12, pp.1819-1829, 2009.

[18] Y. Chen, W. Lin, C. Zhang, Z. Chen, N. Xu, J. Xie, "Intra-and-inter-constraint-based video enhancement based on piecewise tone mapping," *IEEE Trans. Circuits and Systems for Video Technology*, vol. 23, no. 1, pp. 74-82, 2013.

[19] Z. Liu, "A fully automatic system to model faces from a single image," *Microsoft Research*, Technical Report.

[20] S. Wang, X. Gu, H. Qin, "Automatic non-rigid registration of 3D dynamic data for facial expression synthesis and transfer," *CVPR*, pp. 1-8, 2008.

[21] M. Alexa, D. Cohen-Or, D. Levin, "As rigid as possible shape interpolation," *SIGGRAPH*, pp. 157-164, 2000.

[22] Y. Lipman, D. Levin, D. Cohen-Or, "Green coordinates," *SIGGRAPH*, vol. 27, no. 78, 2008.

[23] V. Blanz and T. Vetter T, "A morphable model for the synthesis of 3D faces," Proc. the 26th annual conference on Computer graphics and interactive techniques, pp. 187-194, 1999.

[24] Z. Zhang, Z. Liu, D. Adler, M.F. Cohen, E. Hanson, and Y. Shan, "Robust and rapid generation of animated faces from video images: A model-based modeling approach," *International Journal of Computer Vision*, vol. 58, no. 2, pp. 93-119, 2004.

[25] Y. Sun, X. Wang, and X. Tang, "Deep convolutional network cascade for facial point detection," *IEEE Conf. Computer Vision and Pattern Recognition (CVPR)*, 2013.

[26] C. Cao, Y. Weng, S. Lin, K. Zhou, "3D shape regression for real-time facial animation," *SIGGRAPH*, 2013.